# Representing Extended Finite State Machines for SDL by A Novel Control Model of Discrete Event Systems

*Peng Wang* and *Kai-Yuan Cai*

Department of Automatic Control
Beijing University of Aeronautics and Astronautics
Beijing 100083, China
wp2204@asee.buaa.edu.cn   kycai@buaa.edu.cn

## Abstract

*This paper discusses EFSM for SDL and transforms EFSM into a novel control model of discrete event systems. We firstly propose a control model of discrete event systems, where the event set is made up of several conflicting pairs and control is implemented to select one event of the pair. Then we transform EFSM for SDL to the control model to clarify the control mechanism functioning in SDL flow graphs. This work views the EFSM for SDL in the perspective of supervisory control theory, and this contributes to the field of software cybernetics, which explores the theoretically justified interplay of software and the control.*

## 1. Introduction

Extended Finite State Machines (EFSMs) are widely used in computer science and software engineering [1, 2, 3], especially in the field of program analysis and testing. This model typically simulates flow graphs of software programs, and can be conditionally specified to be the computing model in Specification and Description Language (SDL) [3, 4]. As well known in computer science and engineering, SDL is a language to specify the communication of several processes, where each process is modeled by an EFSM, and the whole communicating system is modeled and computed by CEFSMs, namely Communicating Extended Finite State Machines [5, 6]. This provides one of the backgrounds of our research work.

Another background of this paper refers to the supervisory control theory of discrete event systems, firstly proposed by P. J. Ramadge and W. M. Wonham in 1980s [7, 8]. This theory represents a discrete event system (DES) by an automaton, and control is implemented by another automaton called supervisor. This control structure has been widely accepted and followed up by many other advanced models in this field, such as the partially observable DES, the decentralized control approach and so forth. Nowadays it has formed a theoretical framework, named by RW Framework [9]. Considering the inherent connection of automata and computer science, the supervisory control theory has been introduced into the computer science these years in order to design and analyze the computer program more formally and safely. Some approaches have be explored, and some topics have been discussed with respect to the programming in robotics [10], the software design for the power transformer station [11] as well as developing software by the polynomial dynamic system approach [12].

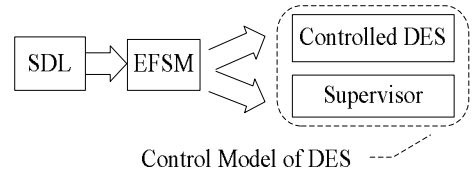

Fig. 1. Represent EFSM for SDL by control model of DES.

In this paper we propose a new control model of discrete event systems on the basis of RW Framework in order to analyze EFSM for the computing model in SDL. This new model is different from that in RW Framework because the new model does not partition the event set by the controllable and uncontrollable subsets. Rather, the event set in the new model is composed of several conflicting pairs. Then the supervisor, which is defined in a similar manner as that in RW Framework, implements control by selecting one event of the conflicting pair. By the new control model, we represent the EFSM for SDL from the perspective of supervisory control theory. The main idea of our research work is illustrated in Fig. 1.

To bridge the gap between EFSM and supervisory control theory, a topic has been introduced as "embedded supervisory control of discrete event systems" in [13]. That is, given a controlled object in terms of a finite automaton and a supervisor in the sense of the RW framework, an equivalent EFMS of a special form can be obtained. In other words, a discrete event system coupled with a supervisor can be transformed to an EFMS in [13], and this also contributes to bridging the same gap. However, we study the topic in the opposite course, that is, we analyze the given EFSM by decomposing it to a novel control model of discrete event systems, and this control model is different from the classic one in RW Framework. Besides this, the EFSM discussed in this paper is more generalized in some aspects than that discussed in [13], and is more applicable to practical problems because it is specified as the computing model of SDL.

One point to be emphasized is, the work of this paper is fundamentally inspired by the thoughts of software cybernetics [14, 15], a new field which studies the theoretically justified interplay between software and the control. And the main result of this paper also contributes to this fields concerning that EFSM, a computing model in computer science, is analyzed from the perspective of supervisory control theory in this paper.

The rest of this paper is organized as follows. Section 2 derives the new control model of discrete event systems. Section 3 introduces EFSM and specifies the general EFSM model to be a special one for SDL. In Section 4 we transform the EFSM for SDL to the new control models via two algorithms. Section 5 concludes this paper and prospects the following research topics.

## 2. Control Model of Discrete Event Systems

This section proposes a new control model of discrete event systems. The new model continues to use the same supervisor/controller structure as that in RW Framework. However, the two models have different partitions on the event sets. That is, the classic model in RW Framework partitions the event set as the controllable and uncontrollable subsets while the new model composes the event set by several conflicting pairs. Based on these pairs supervisory control is implemented by supervisor.

### 2.1 Discrete Event Systems

A discrete event system is modeled by a Mealy finite deterministic automaton
$$G = (Q, \Sigma, Z, \delta, \lambda, q_0)$$
where $Q$ is the state set, $\Sigma$ is the input event set, $Z$ is the output event set, $\delta: Q \times \Sigma \to Q$ is the transition function, $\lambda: Q \times \Sigma \to Z$ is the output function, $q_0 \in Q$ is the initial state. In general, $\delta$ and $\lambda$ are partial functions on the domain, and $\lambda(q, \sigma)$ is defined if and only if $\delta(q, \sigma)$ is defined. Here the transition function can be extended to $\delta: Q \times \Sigma^* \to Q$ as below.

(1) $\delta(q, \varepsilon) = q$
(2) $\delta(q, s\sigma) = \delta(\delta(q, s), \sigma)$

And similarly the output function can be extended to $\lambda: Q \times \Sigma^* \to Z^*$ as below.

(1) $\lambda(q, \varepsilon) = \varepsilon$
(2) $\lambda(q, s\sigma) = \lambda(q, s)\lambda(\delta(q, s), \sigma)$.

It is well known that languages can be used to identify the behaviors of discrete event systems. And as for the Mealy Automaton, it has both input sequences and output sequences. Thus its behavior should naturally be represented by the two sequences combined together. Hence, Given a Mealy Automaton, we define the input language by
$$L_{input}(G) = \{s \in \Sigma^* \mid \delta(q_0, s) \text{ is defined}\},$$
and accordingly define the output language by
$$L_{output}(G) = \{t \in Z^* \mid \exists s \in \Sigma^* \, s.t. \lambda(q_0, s) = t\}.$$

However, it is noted that there exists the situation that more than one input sequence are respondent to a same output sequence. Therefore, it is also necessary to reflect the inherent connection of the input and output sequences in the defined language. So we further give the following definition.
$$L(G) = \{\frac{s}{t} \in (\frac{\Sigma}{Z})^* \mid \delta(q_0, s) \text{ is defined and } \lambda(q_0, s) = t\}$$

And for $(s/t) \in L(G)$, it is easily checked that $s$ and $t$ belong to input and output languages respectively, and $s$ is definitely connected to $t$ by $\lambda(q_0, s) = t$. And therefore this language can completely identify the behavior of the Mealy Automaton. And we also note that the regular expression can be applied here to represent the language especially when the Mealy Automaton is finite.

In this paper the input event set is especially composed of several conflicting pairs. This can be represented by $\Sigma = I \cup \bar{I}$. Consider $I = \{a, b, c\}$ for example, then we have $\bar{I} = \{\bar{a}, \bar{b}, \bar{c}\}$, and this yields the event set $\Sigma = \{a, b, c, \bar{a}, \bar{b}, \bar{c}\}$, where $a$ and $\bar{a}$, $b$ and $\bar{b}$, $c$ and $\bar{c}$ are the given conflicting event pairs. Here we note that this event set can also be represented by $\Sigma = I \times \{true, false\}$, or be interpreted by a relation defined on the event set.

## 2.2 Supervisor

Let $\Sigma = I \cup \bar{I}$, define the control pattern $\gamma$ by
$$\gamma \subseteq \Sigma \text{ and } (\forall \sigma \in \Sigma)(\sigma \in \gamma \Rightarrow \bar{\sigma} \notin \gamma)$$
Then all control patterns compose the set
$$\Gamma = \{\gamma \in 2^\Sigma \mid (\forall \sigma \in \Sigma)(\sigma \in \gamma \Rightarrow \bar{\sigma} \notin \gamma)\}.$$

Event $\sigma$ is said to be enabled by $\gamma$ if $\sigma \in \gamma$; or disabled by $\gamma$ if $\sigma \notin \gamma$. And it is easily checked that control pattern $\gamma$ satisfies $(\forall \sigma \in \Sigma)(\bar{\sigma} \in \gamma \Rightarrow \sigma \notin \gamma)$. Besides, it is also noted that this definition permits the situation that $\bar{\sigma} \notin \gamma \wedge \sigma \notin \gamma$.

The supervisor is a pair $\Phi = (S, \psi)$, where $S = (X, \Sigma, \xi, x_0)$ is a deterministic automaton and $\psi : X \to \Gamma$ is a state feedback map. Couple $\Phi$ to $G$ and this yields the supervised discrete event system by
$$(\Phi / G) = A(X \times Q, \Sigma, Z, \hat{\delta}, \hat{\lambda}, <x_0, q_0>)$$
where the transition function $\hat{\delta}: X \times Q \times \Sigma \to X \times Q$ is defined by
$$\hat{\delta}(x, q, \sigma) = \begin{cases} <\xi(x,\sigma), \delta(q,\sigma)> & \text{if } \sigma \in \psi(x) \text{ and} \\ & \xi(x,\sigma), \delta(q,\sigma) \text{ are defined} \\ \text{undefined} & \text{otherwise} \end{cases}$$
and the output function $\hat{\lambda}: X \times Q \times \Sigma \to Z$ is defined by
$$\hat{\lambda}(x, q, \sigma) = \begin{cases} \lambda(q, \sigma) & \text{if } \hat{\delta}(x, q, \delta) \text{ is defined} \\ \text{undefined} & \text{otherwise} \end{cases}$$

Let $L(\Phi/G)$ denote the language generated by $\Phi/G$. And one point to be emphasized is that, if $\xi(x,\sigma)$ is defined implies $\sigma \in \psi(x)$, then we have
$$\hat{\delta}(x, q, \sigma) \text{ is defined} \Leftrightarrow$$
$$\xi(x, \sigma) \text{ is defined} \wedge \delta(q, \sigma) \text{ is defined}$$
Therefore the supervised system is equivalent to the synchronized product [9] of $S$ and $G$, namely $\Phi/G = S \times G$.

Thereby given a complete supervisor [9] in the form of $\Phi = (S, \psi)$, we can always reduce it via the automaton $S$. Specifically speaking, we firstly reduce $S$ by eliminating $\xi(x, \sigma)$ for $\sigma \notin \psi(x)$, then we further eliminate the inaccessible states and then obtain $S_{\text{modify}}$. Thereby the supervisory control can be equivalently achieved by $G \times S_{\text{modify}}$. Here we note that automaton $S_{\text{modify}}$ corresponds to the supervisor mentioned in [13].

## 3. Extended Finite State Machines for Computing Model of SDL

In this section we firstly introduce the general model of EFSM. Then the general model of EFSM is specilized to be the computing model in SDL. An example is given along with the formal explanation.

### 3.1 General Model of EFSM.

According to [1] we gives the definition of EFSM as follows. An extended finite state machine is structured by a five-tuple
$$EFSM = (Y, I, O, V, T)$$
where $Y$, $I$, $O$ represent the state set, the input event set and the output event set, respectively. $V$ is the variable set, composed of several variables, namely $V = \{v_1, v_2 \cdots v_n\}$. And each variable $v_i$ has its domain, denoted by $dom(v_i)$. Let $\vec{V}$ denote the vector composed by all the variables in $V$, namely $\vec{V} = <v_1, v_2 \cdots v_n>$. Then accordingly we have $dom(\vec{V}) = dom(v_1) \times dom(v_2) \cdots \times dom(v_n)$. $T$ is the transition set. For a transition $t \in T$, $t$ is denoted by
$$t = (y_{t\_src}, y_{t\_dest}, i_t, o_t, P_t, A_t)$$
where $y_{t\_src} \in Y$ and $y_{t\_dest} \in Y$ represent the source state and destination state of the transition, respectively. $i_t \in I$ and $o_t \in O$ represent the input event and the output event, respectively. $P_t(\vec{V})$ is the predicate of the transition, which is defined on the variable set $V$. $A_t(\vec{V})$ is the update function, which updates the values of the variables.

By the definition as above, static structure of EFSM is illustrated. And to make the system dynamic, there are two other points to be mentioned:

(1) The initial condition for EFSM. The initial condition includes the initial state $y_0 \in Y$ and the initial values of the variables, namely $\vec{V}_0 = <v_{10}, v_{20} \cdots v_{n0}>$.
(2) The dynamic rule for EFSM. This rule regulates the dynamic behavior of the system. For the transition $t = (y_{t\_src}, y_{t\_dest}, i_t, o_t, P_t, A_t)$, the rule is explained as follows: when event $i_t$ is imported to state $y_{t\_src}$, if the current value of $\vec{V}$ makes $P_t(\vec{V})$ true, then the transition is enabled, the current state turns to $y_{t\_dest}$, export the event $o_t$ and the value of $\vec{V}$ is updated by $A_t(\vec{V})$; otherwise the transition is disabled, the current state and the value of $\vec{V}$ remain unchanged.

Here we note that an EFSM can be non-deterministic if the current value of $\vec{V}$ enables more than one transition with the one definite input imported. However, this paper only discusses the deterministic EFSM where only one transition is enabled with the one input event at anytime of the system evolvement.

Besides, it is also reasonable to use languages to identify the behavior of the EFSM. And by the similar manners as introduced in Mealy Automata, we can also prefigure the input and output languages as well as the combined language to depict the evolvement behavior of the EFSM model. Here we do not give the formal definition of languages for EFSMs, but corresponding concepts about the languages can be well sensed in the following section.

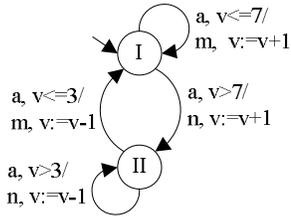

Fig. 2 EFSM for example.

Based on the aforementioned explanation, an example of EFSM is presented in Fig. 2, where the state set is $Y = \{I, II\}$, the input event set is $I = \{a\}$ and the output event set is $O = \{m, n\}$. The variable set is given by $V = \{v\}$, and $dom(v) = \{0,1,2\cdots9\}$. The predicates on variable $v$ are given as Fig. 2 shows, namely $v > 7$, $v \leq 7$, $v > 3$ and $v \leq 3$. The update functions on variable $v$ consist of $v := v+1$ and $v := v-1$. The initial state of EFSM is $I$, and the initial value of $v$ is $v_0 = 0$. Then the EFSM can evolve by the dynamic rule as mentioned above, and this EFSM is definitely deterministic.

## 3.2 EFSM model for SDL.

In this paper, we focus on a special model of EFSM, which functions as the computing model in Specification and Description Language. And we note that SDL specifies the predicate function as Fig. 3 shows.

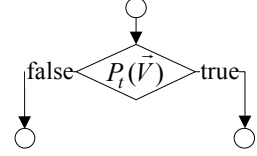

Fig. 3 Predicates in SDL flow graphs.

For an input event, the predicate decides where the system goes by its Boolean values. That is, if the current values of variables make the predicate true, then the system runs to one state; otherwise the system runs to another state.

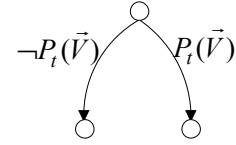

Fig. 4. Predicates in EFSM for SDL.

Therefore the predicate in SDL can be interpreted as a special case of the predicate in general EFSM, as shown in Fig. 4. That is, for the computing model of SDL, the EFSM is characterized by several transition pairs, and each pair consists of two conflicting transitions, of which one is marked by predicate $P_t(\vec{V})$ and the other one is marked by the contrary predicate $\neg P_t(\vec{V})$. Then the transition pair is denoted by $t = (y_{t\_src}, y_{t\_dest}, i_t, o_t, P_t, A_t)$ and $\bar{t} = (y_{t\_src}, y_{t\_dest}, i_t, o_t, \neg P_t, A_t)$.

Therefore the EFSM for the computing model of SDL is formularized by
$$EFSM_{SDL} = (Y, I, O, V, T, y_0, \vec{V}_0)$$
where $Y$, $I$, $O$, $V$ have the same meanings as the counterparts in general EFSM. To incorporate the initial conditions and the dynamic rules in $EFSM_{SDL}$, let $y_0$ and $\vec{V}_0 = <v_{10}, v_{20}\cdots v_{n0}>$ denote the initial state and initial values of variables, and the transition set of the EFSM for SDL is defined by $T: Y \times I \times P(\vec{V}) \times \{true, false\} \to Y \times O \times A(\vec{V})$, where

$P(\vec{V}) \times \{true, false\}$ denotes the conflicting predicates in transition pairs.

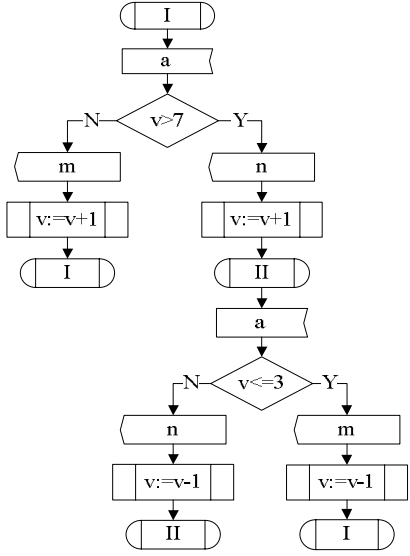

Fig. 5. SDL flow graph.

Then we transform the EFSM given in Fig. 2 to the SDL flow graph in Fig. 5. And the programming code can be further derived from of the SDL flow graph as follows.

```
enum State {I, II}; static State s;
enum Input {a}; enum Output {m,n};
static Input i; static Output o; static int v;
void Initialization()
{ s=I; v=0; }
void Transition (Input i)
{ switch(s)
  { case I:
      switch(i)
        { case a: if(v>7)
              {s=II; o=n; v=v+1;}
            else
              {s=I; o=m; v=v+1;}
          break;
    } break;
      case II:
        switch(i)
        { case a: if(v<=3)
              {s=I; o=m; v=v-1;}
            else
              {s=II; o=n; v=v-1;}
          break;
    } break;
  }
}
```

Here we note that the EFSM for the computing model of SDL must be deterministic because the SDL flow graphs as well as the programming code cannot run ambiguously, but be definitely explained to be deterministic.

## 4. Representing EFSM for SDL by the Control Model of DES

Based on the work of previous sections this section discusses the connection between EFSM and control models of DES. We transform EFSM for SDL to the control model of discrete event systems. Especially, the update function and the predicate together act as a supervisor to control where program flow goes. Here the update function acts as the automaton of the supervisor, and the predicate in EFSM plays the role of state feedback map. The basic idea of this section is shown in Fig. 6.

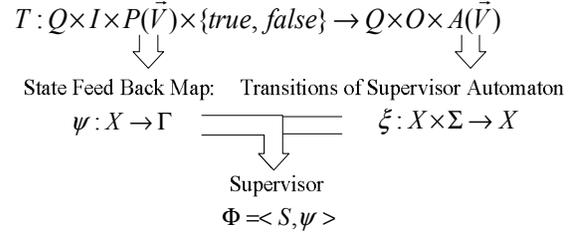

Fig. 6. Extract the supervisor from EFSM for SDL.

This section represents the EFSM for SDL by the control model proposed in Section 1. By this approach we highlight the control functioning in the general program flows, and study the inherent control mechanism in program flow graphs. This work contributes to the field of software cybernetics, which explores the theoretically justified interplay of the software and the control.

In the rest of this section we firstly specify the problem formally, and the solution to the problem is given via two algorithms.

**Problem**: Give an $EFSM_{SDL} = (Y, I, O, V, T, y_0, \vec{V}_0)$, transform this computing model to the control model of DES proposed in Section 1, namely a DES coupled with a supervisor.

### 4.1 Controlled DES

Firstly we extract a controlled DES from EFSM of the computing model in SDL. Consider that the output event set and the output function are defined in

EFSM, the controlled DES is extracted as a Mealy Automaton.

**Algorithm 1**: Give $EFSM_{SDL} = (Y, I, O, V, T, y_0, \vec{V}_0)$. Extract controlled DES $G$.

Step 1. Structure $G$ by $G = (Y, \Sigma, O, \delta, \lambda, y_0)$, where $\Sigma := I \times \{true, false\}$;

Step 2. Let $i \in I$ and $bin \in \{true, false\}$. Construct transition function $\delta : Y \times \Sigma \to Y$ by

$$\delta(y_{src}, <i, bin>) = \begin{cases} y_{dest} & \text{if } T(y_{src}, i, P_t(\vec{V}), bin) \\ & =<y_{dest}, o, A_t(\vec{V})> \\ \text{undefined} & \text{otherwise} \end{cases}$$

Construct the output function $\lambda : Y \times \Sigma \to O$ by

$$\lambda(y_{src}, <i, bin>) = \begin{cases} o & \text{if } T(y_{src}, i, P_t(\vec{V}), bin) \\ & =<y_{dest}, o, A_t(\vec{V})> \\ \text{undefined} & \text{otherwise} \end{cases}$$

Two points should be emphasized here. Firstly we construct the input event set by $\Sigma = I \times \{true, false\}$, and this yields the conflicting event pairs as introduced in Section 1. Secondly, if the transition function $\delta : Y \times \Sigma \to Y$ and the output function $\lambda : Y \times \Sigma \to O$ combine together and additionally consider $\Sigma = I \times \{true, false\}$, then it yields
$$Y \times I \times \{true, false\} \to Y \times O$$
And this corresponds to the transition of $EFSM_{SDL}$ by eliminating $P(\vec{V})$ and $A(\vec{V})$ from $T : Y \times I \times P(\vec{V}) \times \{true, false\} \to Y \times O \times A(\vec{V})$.

Recalling the example given in Section 3, we extract the Mealy Automaton to be the controlled DES as shown in Fig. 7. And the conflicting event pair is obtained by $<a, true>$ and $<a, false>$.

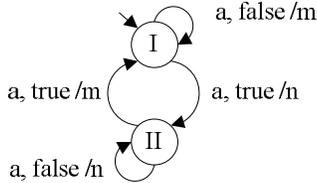

Fig. 7. Controlled DES extracted from EFSM.

### 4.2 Supervisor

Secondly we extract the supervisor from EFSM. To obtain the supervisor, two steps should be taken: (1) construct automaton $S$ by the update function $A(\vec{V})$; (2) construct state feedback map $\psi$ by the predicate $P(\vec{V})$.

**Algorithm 2**: Give $EFSM_{SDL} = (Y, I, O, V, T, y_0, \vec{V}_0)$. Extract supervisor $\Phi = (S, \psi)$.

Step 1. Structure automaton $S = (X, \Sigma, \xi, x_0)$, where
$X := dom(\vec{V}) \times Y$; $\Sigma := I \times \{true, false\}$;
$x_0 :=<\vec{V}_0, y_0>$.

Step 2. Let $bin \in \{true, false\}$. Construct transition $\xi : X \times \Sigma \to X$ as follows:
consider $\xi(<\vec{V}, y_{src}>, <i, bin>)$, if $T(y_{src}, i, P_t(\vec{V}), bin) =<y_{dest}, o, A_t(\vec{V})>$ is defined and $A_t(\vec{V}) \in dom(\vec{V})$, then $\xi(<\vec{V}, y_{src}>, <i, bin>) =<A_t(\vec{V}), y_{dest}>$; otherwise $\xi(<\vec{V}, y_{src}>, <i, bin>)$ is undefined.

Step 3. Let $bin \in \{true, false\}$. Let $\text{proj}_{\vec{V}}(x)$ and $\text{proj}_y(x)$ denote the segments $\vec{V}$ and $y$ subject to state $x$. Namely, given $x =<\vec{V}, y>=<v_1, v_2 \cdots v_n, y>$, we have $\text{proj}_{\vec{V}}(x) = \vec{V} =<v_1, v_2 \cdots v_n>$ and $\text{proj}_y(x) = y$.

Construct state feedback map $\psi : X \to \Gamma$ as follows: $<i, bin> \in \psi(x)$ if and only if $T(\text{proj}_y(x), i, P_t(\vec{V}), bin)$ is defined and $P_t(\vec{V})|_{\vec{V}=\text{proj}_{\vec{V}}(x)} = bin$.

Considering the classical logic system, $\vec{V} = \text{proj}_{\vec{V}}(x)$ can only make $P_t(\vec{V})$ to be true or false. And note that only deterministic EFSMs are discussed, it is easily checked that the state feedback map follows the definition given in Section 1. That is, $<i, true> \in \psi(x)$ implies $<i, false> \notin \psi(x)$.

Recalling the example presented in Section 3, we construct the state set by $X = dom(v) \times \{I, II\}$. The initial state for automaton $S$ is obtained by $x_0 = 0I$. And the event set for automaton $S$ is $\Sigma = \{<a, true>, <a, false>\}$.

Apply Step 2 in Algorithm 2, and we construct the transition function of automaton $S$. And automaton $S$ is illustrated in Fig. 8.

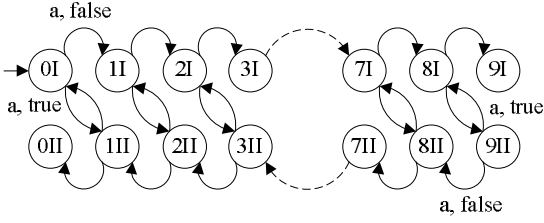

Fig. 8. Automaton of the supervisor.

Apply Step 3 in Algorithm 2, and we construct the state feedback map by the predicates in the EFSM. And the supervisor is finally obtained as shown below.

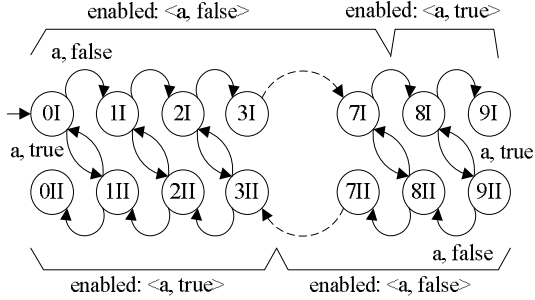

Fig. 9. Supervisor extracted from EFSM.

According to Section 1 the supervisory control implemented by $\Phi = (S, \psi)$ can be equivalently implemented by an automaton $S_{\text{modify}}$. Then the supervised system can be achieved by the synchronized product of the controlled DES and the automaton, namely $G \times S_{\text{modify}}$. Here we illustrate $S_{\text{modify}}$ in Fig. 10.

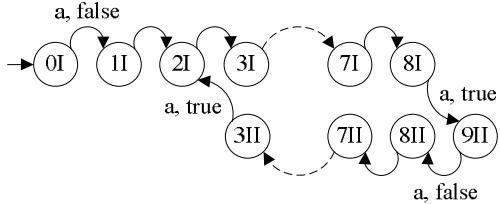

Fig. 10. Supervisor modified.

After introducing the algorithms we give explanation as follows. It is noted that the evolvement of state-transition systems can be represented by the event sequences, and both of the EFSM and the control model of DES have the input sequences and output sequences, where each input sequence corresponds to an output sequence definitely. So the question is whether the algorithms keep the sequences fixed between the two models. The answer is affirmative. Although the input event set of the control model derived is $I \times \{true, false\}$, we can actually extract the first sub-element from it, and this makes the two models comparable with respect to input events. And thereby it is checked that the two models have the same set of the input sequences as well as the corresponding the output sequences, that is, the two models follow the same disciplines during evolvement.

Consider the forgoing example, and we illustrate the input and output sequences of the control model by the diagram as below.

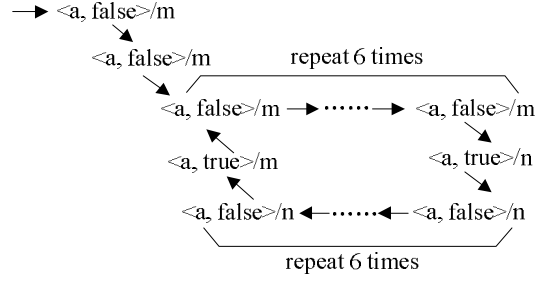

Fig. 11. The input and output sequences of the control model.

Here regular expressions can be derived to represent the sequences as shown in Fig. 11. And we extract the first sub-elements of the control model's events. Then $<a, true>$ and $<a, false>$ are simplified as $a$. And if we mark state $I$ and $v = 2$, Then the input language is formulated by $a^2(a^{14})^*$, and the output language is given by $m^2(m^6 n^7 m)^*$. Consequently we can formulate the combined sequences by the language as below.

$$(\frac{a}{m})^2((\frac{a}{m})^6(\frac{a}{n})^7 \frac{a}{m})^*$$

Consider the EFSM presented in Section 3, we illustrate its input and output sequences by the diagram in Fig. 12.

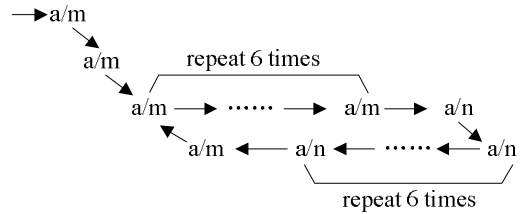

Fig. 12. The input and output sequences of EFSM.

Then it is easily checked that the EFSM above has the same input and output sequences as the derived control model. Specifically speaking, from the diagram in Fig. 12 we can exactly deduce the same input and output languages as the derived control model. Furthermore, if we also mark state $I$ and $v = 2$, then the same regular expression can be formulated to denote the combined sequence as the same as that of the control model.

## 5. Concluding Remarks

This paper refers to two research topics, the control model of DES and the EFSM for SDL, and bridges the gap between the two topics via the given algorithms. We firstly propose a novel control model of DES, which is characterized by selecting one event of the conflicting event pair. Then we focus on EFSM, and specify the general EFSM to be the computing model for SDL. Then the connection between these two models is studied and the algorithms are given to transform the EFSM for SDL to the control model of DES.

The work of this paper highlights the control mechanism functioning in flow graphs of SDL, and contributes to the field of software cybernetics, which explores the theoretically justified interplay between software and the control. Based on the work of this paper, some follow-up topics can be approached. And one topic refers to Communicating Extended Finite State Machines, which functions as the communicating model in SDL. We can analyze CEFSM based on the work of this paper and further propose a novel approach to verify the communicating system modeled by SDL from the perspective of supervisory control theory.